\documentclass[a4paper,12pt]{article}
\usepackage{cite}
\usepackage{amssymb}
\usepackage{amsmath}
\usepackage{amsfonts}
\usepackage{amsthm}
\usepackage{mathtools}
    \DeclareMathOperator{\hc}{h.c.}
\usepackage[utf8]{inputenc}
\usepackage{color}
\usepackage{graphicx}
\usepackage{color}
\usepackage[active]{srcltx}

\usepackage{here}
\usepackage{cite}
\usepackage{soul}
\usepackage[affil-it]{authblk}

\usepackage{titlesec}
\titleformat{\subsection}
       {\normalfont\fontsize{13}{17}\itshape}{\thesubsection}{1em}{}

\input paperdef

\usepackage{url}
\begin{document}

\thispagestyle{empty}

\title{
Reduction of Couplings in the Type-II 2HDM}
\date{}
\author{M. A. May Pech$^{1}$\thanks{email: miguel.maypech@gmail.com}~, 
M. Mondrag\'on$^{1}$\thanks{email: myriam@fisica.unam.mx}~, 
G. Patellis$^{2}$\thanks{email: grigorios.patellis@tecnico.ulisboa.pt}~
and G. Zoupanos$^{3,4,5,6}$\thanks{email: george.zoupanos@cern.ch}\\
{\small
$^1$Instituto de F\'{\i}sica, Universidad Nacional Aut\'onoma de M\'exico,\\
A.P. 20-364, CDMX 01000 M\'exico\\
$^2$ Centro de Física Teórica de Partículas - CFTP, Departamento de Física,\\ Instituto Superior Técnico, Universidade de Lisboa,\\
Avenida Rovisco Pais 1, 1049-001 Lisboa, Portugal \\
$^3$ Physics Department,   National Technical University, 157 80 Zografou, Athens, Greece\\
$^4$ Institut für Theoretische Physik der Universität Heidelberg,\\
Philosophenweg 16, 69120 Heidelberg, Germany\\
$^5$ Max-Planck Institut f\"ur Physik, F\"ohringer Ring 6, D-80805
  M\"unchen, Germany \\
$^6$ Theoretical Physics Department, CERN, Geneva, Switzerland \\
}
}

{\let\newpage\relax\maketitle}

\begin{abstract}

The idea of \textit{reduction of couplings} consists in the search for relations between seemingly independent couplings of a renormalizable theory that are renormalization group invariant. In this article, we  demonstrate the existence of such 1-loop relations among the top Yukawa, the Higgs quartic and the gauge colour couplings of the Type-II Two Higgs Doublet Model at a high-energy boundary. The phenomenological viability of the reduced theory suggests the value of $\tan\beta$ and  the scale in which new physics may appear.
\end{abstract}


\section{Introduction}\label{intro}

An essential~direction of the last decades in theoretical particle physics is to understand the free  parameters of the Standard Model (SM) in terms of a  few
fundamental ones, i.e. to achieve a {\it reduction of couplings} (RoC) \cite{book}. However, despite the numerous successes of the SM regarding the description of elementary particles and the interactions among them, there is significantly less progress when it comes to the freedom in the parameter space. The problem of the large number of arbitrary parameters is deeply related to the infinities that emerge at the quantum~level. While renormalization succeeds in removing those infinities, it only does so at the cost of introducing counter terms, which leaves the `cured' parameters free to be fixed by the experiment.

Although the success of the SM is undisputed, it is a widespread belief that it is ultimately the low energy limit of~a (more) fundamental theory. Under this light, one of the most popular and efficient ways to  reduce that freedom in  parameter space is to introduce a symmetry. A well known example of such an~idea are the Grand Unified~Theories (GUTs)\cite{Pati:1973rp,Georgi:1974sy,Georgi:1974yf,Fritzsch:1974nn,Gursey:1975ki,Achiman:1978vg}. Within the GUT framework, gauge couplings are related and one 
can even have a unified Yukawa sector. Unfortunately, beyond the minimal $SU(5)$, which was experimentally ruled out a long time ago, theories based on larger groups  give rise to new complications regarding the number of free parameters since new degrees of freedom are necessary (i.e the channels~of breaking the symmetry).

The RoC method was proposed as an alternative, systematic way to look for relations among seemingly~unrelated parameters \cite{Zimmermann:1984sx,Oehme:1984yy,Oehme:1985jy} (see~also \cite{Ma:1977hf,Chang:1974bv,Nandi:1978fw}).
This technique~reduces the number of independent parameters of a theory by relating either all (in its original version) or a number of parameters to a single coupling, which is often called `primary coupling'. For this approach to be considered systematic, two conditions should hold. First, both the original and the reduced theory should be renormalizable. Second, the relations among the various parameters involved should be renormalization group invariant (RGI). 

This idea was of course first applied on the SM almost four decades ago \cite{Kubo:1985up,Kubo:1988zu} and, while at the time it produced promising predictions for the top quark and Higgs boson masses, their following respective experimental discoveries ruled them out as too light. However, this work opened the way to a number of theories that extend the SM and are based on the  concept of the reduction of couplings, which had significant predictive power and were \cite{Kapetanakis:1992vx,Kubo:1994bj,Mondragon:1993tw,Kobayashi:1997qx,Heinemeyer:2007tz} or continue to be \cite{Heinemeyer:2020ftk,Heinemeyer:2020nzi} successful.

In the present work we choose a minimal extension of the SM, namely the well known Two Higgs Doublet Model (2HDM) \cite{Lee:1973iz} and, more specifically, its Type-II version (this is just a convenience choice, since similar procedures as the one here described can be applied to all versions). In particular, we use a version of the RoC technique first introduced in \cite{Heinemeyer:2017gsv} in search of a boundary scale at which we have New Physics. The elegance of this approach is that only a second Higgs doublet is needed in order to fit the reduced model with the current experimental constraints and pinpoint the scale above which new field content and/or symmetries could come into effect.

\section{Reduction of couplings basics}\label{roc}

A brief description of the  basic idea of \textit{reduction of couplings} is  first in order, as  it was introduced in \cite{Zimmermann:1984sx} and consequently expanded over the next decades. The goal is to express the couplings ~of a theory that~are considered free in terms of one parameter, that is considered to be more fundamental, called the primary coupling. The basic idea is to search for renormalization group invariant (RGI) relations among parameters that reduce the degree  of arbitrariness of the parameter space.

Such relations are in general of the form $F (g_1,\cdots,g_A) ~=~\mbox{const.}$ for $A$ number of parameters, which should satisfy the partial differential equation (PDE):
\beq
\mu\,\frac{d F}{d \mu} = {\vec \nabla}F\cdot {\vec \beta} ~=~
\sum_{a=1}^{A}
\,\beta_{a}\,\frac{\partial F}{\partial g_{a}}~=~0~,
\eeq
where $\beta_a$ are the~$\beta$-functions of each coupling $g_a$, respectively, in order for  $F$ to be RGI. This PDE is equivalent to the set of ordinary differential equations below, which are called reduction equations (REs) \cite{Zimmermann:1984sx,Oehme:1984yy,Oehme:1985jy},
\beq
\beta_{g} \,\frac{d g_{a}}{d g} =\beta_{a}~,~a=1,\cdots,A-1,
\label{redeq}
\eeq
where now $g$ and  $\beta_{g}$ are the  primary
coupling and its respective $\beta$-function. 
There are -maximally- $A-1$ independent
RGI constraints in the  $A$-dimensional space of parameters
imposed by the  $F_a$'s, thus  one could in principle
express all parameters in terms of one
primary coupling, $g$.

However, the general solutions of the REs contain as many integration constants as the number of the equations themselves. Thus, so far we have just~traded an integration constant for each coupling and these general solutions cannot be considered to have reduced the freedom of the parameter space. The  crucial requirement is that the REs admit power series solutions:
\beq
g_{a} = \sum_{n}\rho_{a}^{(n)}\,g^{2n(+1)}~,
\label{powerser}
\eeq
which~preserve~perturbative~renormalizability. Remarkably, the  uniqueness of these power series solutions can be already decided at 1-loop level
\cite{Zimmermann:1984sx,Oehme:1984yy,Oehme:1985jy}.

The possibility of a \textit{complete} reduction of couplings described above is without doubt very attractive, as the completely reduced theory features only one independent coupling. However, in many cases this has been proven to be unrealistic. Therefore, fewer RGI  constraints are often imposed, leading to a \textit{partial} reduction \cite{Kubo:1985up,Kubo:1988zu}  of the parameter space.

\section{Notation and parameters of the 2HDM}\label{2hdm}

For the two Higgs doublets $\Phi_1,\Phi_2$ the most general renormalizable scalar potential can be written as \cite{Wu:1994ja,Davidson:2005cw,Branco:2011iw}:
\begin{align}\label{potential}
    V_h=&~~~m_{11}^2\Phi_1^{\dagger}\Phi_1+m_{2}^2\Phi_2^{\dagger}\Phi_2-\Big(m_{12}^2\Phi_1^{\dagger}\Phi_2+ \hc \Big)\nonumber\\
    &+\frac{1}{2}\lambda_1\Big(\Phi_1^{\dagger}\Phi_1\Big)^2+\frac{1}{2}\lambda_2\Big(\Phi_2^{\dagger}\Phi_2\Big)^2+\lambda_3\Big(\Phi_1^{\dagger}\Phi_1\Big)\Big(\Phi_2^{\dagger}\Phi_2\Big)+\lambda_4\Big(\Phi_1^{\dagger}\Phi_2\Big)\Big(\Phi_2^{\dagger}\Phi_1\Big)\nonumber\\
    &+\Bigg[\frac{1}{2}\lambda_5\Big(\Phi_1^{\dagger}\Phi_2\Big)^2+\lambda_6\Big(\Phi_1^{\dagger}\Phi_1\Big)\Big(\Phi_1^{\dagger}\Phi_2\Big)+\lambda_7\Big(\Phi_2^{\dagger}\Phi_2\Big)\Big(\Phi_1^{\dagger}\Phi_2\Big)+ \hc\Bigg],
\end{align}
where $m_{11}^2,m_{22}^2$ and $\lambda_{1,2,3,4}$ are always real, while $m_{12}^2$ and $\lambda_{5,6,7}$ are in general complex. Since in this work we want to demonstrate the simplest possible application of RoC on a 2HDM, we choose to consider all of the above-mentioned parameters to be real.

The discrete symmetries introduced in the context of the Type-II scenario (in which $u_R^i$ couple with $\Phi_2$, while $d_R^i$ and $e_R^i$ couple with $\Phi_1$) ensure that 
\begin{equation}\label{const1}
\lambda_6=\lambda_7~~,
\end{equation}
while in order to conserve the electric charge one needs:
\begin{equation}\label{const2}
\lambda_4<0~~.
\end{equation}
Furthermore, the potential is bounded from below if \cite{hlr85,Branco:2011iw,ivanov17}
\begin{align}\label{const3}
\lambda_1>0~~, &&
\lambda_2>0~~, &&
\sqrt{\lambda_1\lambda_2} + \lambda_3  + \lambda_4 - |\lambda_5| > 0 ~~.
\end{align}

\section{A first attempt of reduction}\label{firstattempt}

Let us now proceed with the reduction of the parameters of the model. As in past reduced models, the best candidate for `primary' coupling is the strong coupling, $g_s$. A complete reduction is not realistic, so we focus on the third fermionic generation Yukawa couplings. Furthermore, reducing in favour of a dimensionless parameter only works for dimensionless parameters, so $m_{11}^2, m_{22}^2$ and $m_{12}^2$ will remain free. 

The gauge couplings $g$ and $g'$ of the $SU(2)$ and $U(1)$ gauge groups, respectively,  will not be considered at the first stage of the reduction, but will be treated as corrections. Since the  bottom quark and tau lepton Yukawa couplings are much smaller than the top Yukawa coupling, we do not take them into account in the following work for simplicity. However, they can be straightforwardly incorporated into the following reduction scheme in future studies of the model. \\

First, we have to specify the 1-loop renormalization group equations (RGEs), which were given (for a general gauge theory and for the specific case of two scalar doublets) in \cite{Cheng:1973nv,Jones:1981we,Machacek:1981ic,Haber:1993an,Grimus:2004yh,Ferreira:2010xe}. For  coherency, we follow the notation of \cite{Branco:2011iw}. The gauge $\beta$-functions for the model are given by:
\begin{align}
    \mathcal{D}g_s~=&-7g_s^3~\equiv~\beta_3\label{g3}\\
    \mathcal{D}g~=&-3g^3~\equiv~\beta_2\label{g2}\\
    \mathcal{D}g'~=&~~~7g'^3~\equiv~\beta_1~,\label{g1}
\end{align}
where $\mathcal{D}$ is the dimensionless differential operator $16\pi^2\mu(d/d\mu)$. The  top Yukawa $\beta$-function  (with the omission of $y_b$ and $y_\tau$) is:
\begin{equation}
    \mathcal{D}y_t~=~\beta_t~=~\beta_{t_0}+\beta_{t_c}~,\label{yt}
\end{equation}
where
\begin{align}
    \beta_{t_0}~=&~\Big(\frac{9}{2}y_t^2-8g_s^2\Big)y_t\label{yt_0}\\
    \beta_{t_c}~=&~\Big(-\frac{9}{4}g^2-\frac{17}{12}g'^2\Big)y_t~,\label{yt_c}
\end{align}
and it is understood that $\beta_{t_0}$ is the top Yukawa $\beta$-function without the $g,g'$ contributions, which are notated  as $\beta_{t_c}$. The $\lambda_i$ $\beta$-functions -without the bottom and tau contributions- are  given by:
\begin{equation}\label{li}
   \mathcal{D}\lambda_i~=~\beta_{\lambda_i}~=~\beta_{{\lambda_i}_0}+\beta_{{\lambda_i}_c}~,
\end{equation}
where again $\beta_{{\lambda_i}_0}$ are the $\lambda_i$ $\beta$-functions without the $g,g'$ contributions $\beta_{{\lambda_i}_c}$ and are given as
\begin{align}
\beta_{{\lambda_1}_0}~=&~12\lambda_1^2+4\lambda_3^3+4\lambda_3\lambda_4+2\lambda_4^2+2\lambda_5^2+24\lambda_6^2\label{l1_0}\\
\beta_{{\lambda_2}_0}~=&~12\lambda_2^2+4\lambda_3^3+4\lambda_3\lambda_4+2\lambda_4^2+2\lambda_5^2+24\lambda_7^2+12\lambda_2\lambda_t^2-12y_t^4\label{l2_0}\\
\beta_{{\lambda_3}_0}~=&~(\lambda_1+\lambda_2)(6\lambda_3+2\lambda_4)+4\lambda_3^2+2\lambda_4^2+2\lambda_5^2+4(\lambda_6^2+\lambda_7^2)+16\lambda_6\lambda_7+6\lambda_3y_t^2\label{l3_0}\\
\beta_{{\lambda_4}_0}~=&~2(\lambda_1+\lambda_2)\lambda_4+8\lambda_3\lambda_4+4\lambda_4^2+8\lambda_5^2+10(\lambda_6^2+\lambda_7^2)+4\lambda_6\lambda_7+6\lambda_4y_t^2\label{l4_0}\\
\beta_{{\lambda_5}_0}~=&~\lambda_5(2\lambda_1+2\lambda_2+8\lambda_3+12\lambda_4)+10(\lambda_6^2+\lambda_7^2)+4\lambda_6\lambda_7+6\lambda_5y_t^2\label{l5_0}\\
\beta_{{\lambda_6}_0}~=&~(12\lambda_1+6\lambda_3+8\lambda_4)\lambda_6+(6\lambda_3+4\lambda_4)\lambda_7+10\lambda_5\lambda_6+2\lambda_5\lambda_7+9\lambda_6y_t^2\label{l6_0}\\
\beta_{{\lambda_7}_0}~=&~(12\lambda_1+6\lambda_3+8\lambda_4)\lambda_7+(6\lambda_3+4\lambda_4)\lambda_6+10\lambda_5\lambda_7+2\lambda_5\lambda_6+9\lambda_7y_t^2\label{l7_0}
\end{align}
and 
\begin{align}
    \beta_{{\lambda_1}_c}~=&~\frac{3}{4}(3g^4+g'^4+2g^2g'^2)-3\lambda_1(3g^2+g'^2)\label{l1_c}\\
    \beta_{{\lambda_2}_c}~=&~\frac{3}{4}(3g^4+g'^4+2g^2g'^2)-3\lambda_2(3g^2+g'^2)\label{l2_c}\\
    \beta_{{\lambda_3}_c}~=&~\frac{3}{4}(3g^4+g'^4-2g^2g'^2)-3\lambda_3(3g^2+g'^2)\label{l3_c}\\
    \beta_{{\lambda_4}_c}~=&~3g^2g'^2-3\lambda_4(3g^2+g'^2)\label{l4_c}\\
    \beta_{{\lambda_5}_c}~=&~-3\lambda_5(3g^2+g'^2)\label{l5_c}\\
    \beta_{{\lambda_6}_c}~=&~-3\lambda_6(3g^2+g'^2)\label{l6_c}\\
    \beta_{{\lambda_7}_c}~=&~-3\lambda_7(3g^2+g'^2)\label{l7_c}~,
\end{align}
Proceeding with the reduction of $\lambda_i$ and $y_t$ w.r.t. the primary coupling, $g_s$ and temporarily `switching off' the other two gauge couplings, the power series solutions of \refeq{powerser} will be:
\begin{align}
    y_t~=&~p_tg_s\label{pst0}\\
    \lambda_i~=&~p_ig_s^2\label{psl0}~.
\end{align}
Substituting the solutions into the REs,
\begin{align}
    \beta_3\frac{dy_t}{dg_s}~=&~\beta_{t_0}\label{REt0}\\
    \beta_3\frac{d\lambda_i}{dg_s}~=&~\beta_{{\lambda_i}_0}~,\label{REl0}
\end{align}
we get sets of $p_i,\,p_t$ that depend on $\sin\beta$ and are RGI. Sadly, these solutions do not satisfy \refeq{const2} and \refeq{const3}  and, more importantly, for any choice of $\tan\beta$ (which is the well known ratio of the two Higgs vacuum expectation values - vevs), the top quark pole mass fails to go above $100$ GeV. 

One may proceed with taking into account the corrections that come from the other two gauge couplings, in hope that their contributions may ameliorate the above results. Now the full power series solutions of \refeq{powerser} will be:
\begin{align}
    y_t~=&~p_tg_s+q_tg+r_tg'\label{pst}\\
    \lambda_i~=&~p_ig_s^2+q_ig^2+r_ig'^2\label{psl}~,
\end{align}
where $p_t,p_i$ are known from the above procedure and the corresponding REs will be
\begin{align}
    \beta_3\frac{dy_t}{dg_s}~=&~\beta_t\label{REt}\\
    \beta_3\frac{d\lambda_i}{dg_s}~=&~\beta_{\lambda_i}~.\label{REl}
\end{align}
In order for \refeqs{REt}-(\ref{REl}) to be solved w.r.t. $q_t,q_i,r_t,r_i$, one needs a further condition such as \cite{Heinemeyer:2017gsv}
\begin{equation}
    \mathcal{D}(q_a g)\sim 0~~~,~~~\mathcal{D}(r_a g')\sim0~,\label{condition}
\end{equation}
where $a=t,1,...,7$. However,  these conditions only hold for $\mu\geq 10^7$ GeV and are
not RGI, thus we  cannot have a successful reduction of the theory that way.

An  extended discussion about `traditional' (partial) RoC method applied directly to the 2HDM, can be found in \cite{denner90}, where a reduction is performed for the Yukawa and Higgs self-couplings in terms of the strong coupling. Using the bottom quark mass as input the Higgs boson and top quark masses are predicted (with values which are now ruled out by experiment), and also values for the lepton masses and the extra Higgs scalars were found, 
but already in contradiction with the experimental results at the time. 
More recently, a similar study of the RoC method specifically applied to the four types of 2HDM with Natural Flavour Conservation with updated data was performed in \cite{thesisMA}, with similar, albeit not identical, results.

\section{A realistic approach to reduction}\label{realistic}

Since the `traditional' partial RoC method proved to be too restrictive, the next step is to try a reduction at a boundary scale, along the lines of \cite{Heinemeyer:2017gsv} (also explained in \cite{Heinemeyer:2019vbc}). The idea is simple: we solve the REs of \refeqs{REt}-(\ref{REl}) at one specific scale,  called the boundary scale $M_{\text{bdry}}$, above which a covering theory is assumed, which is supposed to make these solutions RGI. Below $M_{\text{bdry}}$ we run the usual 2HDM RGEs, using the reduction solutions as boundary conditions. Thus, using only the experimental values of the gauge couplings and by fixing the $\tan\beta$ value, we obtain the top quark pole  mass and all $\lambda_i$s at the EW scale and, fixing the mass parameters of the scalar potential, we also obtain the light Higgs boson mass. 

The conditions of \refeq{condition} demand that $M_{\text{bdry}}\geq 10^7$ GeV. However, since we want to treat $g,g'$ as corrections, we need their values not to be comparable to $g_s$ at the boundary scale. Running the 2HDM gauge RGEs of \refeqs{g3}-(\ref{g1}) it becomes clear that, while $g'$ continues to be much smaller than the other gauge couplings until very high energies, the weak coupling starts `dangerously' approaching $g_s$ around $\mu\sim 10^8$ GeV. This naturally restricts  the boundary scale at
\begin{equation}
    M_{\text{bdry}}\sim 10^7~\text{GeV ,}
\end{equation}
which is the value used from now on. The reduction of $y_t$ to  $g_s$ does not involve any of the Higgs potential parameters and can be performed independently. 
As such, we reduce the top Yukawa in favour of the strong coupling, first without the contributions of $g,g'$, as above. Then, in order to solve the RE of \refeq{REt} at the boundary $M_{\text{bdry}}$, we use the values $g(M_{\text{bdry}})$ and $g'(M_{\text{bdry}})$ that we get from \refeqs{g2}-(\ref{g1}) using their experimental values at $M_Z$. From three sets of possible reduction solutions at the boundary scale, the only one that  can lead to a phenomenologically viable top mass is
\begin{equation}
    y_t~=~0.471g_s-0.119g+1.228g'~.\label{pstMB}
\end{equation}
Using the above relation as boundary condition to the top Yukawa RGE, we obtain the EW scale top Yukawa value. In order to satisfy the experimental constraint of \cite{ParticleDataGroup:2022pth},
\begin{equation}
    m_t=(172.69\pm 0.30)~\text{GeV}~,\label{topexp}
\end{equation}
allowing a theoretical uncertainty of $1$ GeV, the ratio of the two Higgs vevs has to be:
\begin{equation}
    \tan\beta~=~2.2\pm 0.1 ~.\label{tanb}
\end{equation}
Now, with all the above information, we can perform the same reduction to the full system of $g_s$, $y_t$ and $\lambda_i$  at $M_{\text{\text{bdry}}}$, including the corrections of $g,g'$. We get a large number of possible solutions, not all of which are phenomenologically viable. Indeed, once we impose  the conditions of \refeqs{const1}-(\ref{const3}), there are only five sets of reduction solutions that confirm the light Higgs boson mass measurement \cite{ParticleDataGroup:2022pth},
\begin{equation}
m_h^{\text{exp}}=(125.25\pm 0.17)~\text{GeV}\,. \label{higgsexp}
\end{equation}
We have estimated that our theoretical calculations have a $10$ GeV uncertainty, due to threshold corrections and higher order contributions. For the calculation  of the light Higgs mass we have chosen appropriate mass parameters of the Higgs potential such that the mass of the CP odd scalar, $m_A$, is  $800$ GeV, which is allowed by recent LHC searches \cite{Wang:2022yhm}. However, there is significant freedom in this parameter, since a variation of $\pm 400$ GeV in $m_A$ gives a very small change in the value of $m_h$, which is covered by the theoretical uncertainty.
The five reduction solutions are shown in \refta{tab:solutions}, while the results for their respective light Higgs masses are given in \refta{tab:higgsmass}. It is obvious from  \refta{tab:solutions} and the $\lambda_i$ $\beta$-functions that every solution has $\lambda_{6,7}=0$, while $\lambda_5$  vanishes for the latter two solutions. 

\begin{table}
\renewcommand{\arraystretch}{1.5}
\centering
\begin{tabular}{|c|rrrrrrrr|}
\hline
\# & $p_t$ & $p_1$ & $p_2$ & $p_3$ & $p_4$ & $p_5$ & $p_6$ &
  $p_7$\\
\hline
SET1 & 0.471 & -1.167 & -1.424 & 0 & 0 & 0 & 0 & 0  \\
SET2 & 0.471 & -1.167 & -1.424 & 0 & 0 & 0 & 0 & 0  \\
SET3 & 0.471 & -0.748 & -1.162 & -0.970 & 0 & 0 & 0 & 0  \\
SET4 & 0.471 & -0.748 & -1.162 & -0.970 & 0 & 0 & 0 & 0  \\
SET5 & 0.471 & -0.748 & -1.162 & -0.970 & 0 & 0 & 0 & 0  \\

\hline

\hline
\# & $q_t$ & $q_1$ & $q_2$ & $q_3$ & $q_4$ & $q_5$ & $q_6$ &
  $q_7$\\
\hline
SET1 & -0.119 & 4.183 & 4.687 & 0.170 & -0.749 & 0 & 0 & 0   \\
SET2 & -0.119 & 4.183 & 4.687 & -0.579 & -0.749 & 0 & 0 & 0  \\
SET3 & -0.119 & 0.939 & 3.483 & 4.327 & -1.403 & -1.403 & 0 & 0  \\
SET4 & -0.119 & 2.738 & 3.772 & 3.347 & 0 & 0 & 0 & 0  \\
SET5 & -0.119 & 2.738 & 3.772 & 3.347 & 0 & 0 & 0 & 0  \\

\hline

\hline
\# & $r_t$ & $r_1$ & $r_2$ & $r_3$ & $r_4$ & $r_5$ & $r_6$ &
  $r_7$\\
\hline
SET1 & 1.228 & 0.189 & -3.824 & 1.564 & -2.463 & 0 & 0 & 0  \\
SET2 & 1.228 & 1.123 & -3.275 & 5.623 & -3.774 & -3.774 & 0 & 0  \\
SET3 & 1.228 & 5.168 & -3.849 & 9.539 & -4.847 & -4.847 & 0 & 0 \\
SET4 & 1.228 & 0.247 & -4.148 & 7.668 & -8.187 & 0 & 0 & 0  \\
SET5 & 1.228 & 5.855 & -4.110 & 6.094 & -7.863 & 0 & 0 & 0  \\

\hline
\end{tabular}

\caption{\textit{
The sets of solutions for $M_{\text{\text{bdry}}}=10^7$ GeV and $\tan\beta=2.2$ that satisfy the conditions of \refeqs{const1}-(\ref{const3}) and yield a light Higgs boson mass within $10$ GeV of  \refeq{higgsexp}.}}\label{tab:solutions}
\renewcommand{\arraystretch}{1.0}
\end{table}

\begin{table}[htb!]
\renewcommand{\arraystretch}{1.5}
\centering
\begin{tabular}{|c|r|}
\hline
\# & $m_h$ (GeV) \\
\hline
SET1 & 124.55 \\
SET2 & 118.06 \\
SET3 & 116.12 \\
SET4 & 121.70 \\
SET5 & 121.36 \\
\hline
\end{tabular}

\caption{\textit{
The light Higgs boson mass (in GeV) for each set of solutions of \refta{tab:solutions}.}}\label{tab:higgsmass}
\renewcommand{\arraystretch}{1.0}
\end{table}

This is the first case in which the RoC method is applied on a non-supersymmetric model and successfully fits the experimentally observed values for both the top quark mass and the (light) Higgs boson mass. It assumes new physics at a specific energy scale -either a covering symmetry or just new field content, or both- with RGI relations among parameters. It is only a simplified example of what the method is capable of, as it can be expanded to either be more restrictive  or include other phenomena as well. For  example, a 2-loop analysis of the above can rule out some of the solutions, while it can also be applied on a complex Higgs potential. In the latter case, it can result in a realistic description of explicit or spontaneous CP violation with minimal input, and it may single out one of 
the six symmetries  of the Higgs potential  \cite{Ferreira:2009wh,Branco:2011iw} as the one that naturally occurs from a reduced theory. Lastly, although the extension of the SM with one scalar doublet is one of  the simplest and most intuitive ways to tackle questions the SM is unable to, the RoC technique can be applied in more field-rich models (or models  with larger symmetries like in \cite{Kubo:1994bj,Kapetanakis:1992vx,Mondragon:1993tw,Heinemeyer:2007tz,Heinemeyer:2020ftk,Heinemeyer:2020nzi}). The natural continuation of the present work under this perspective is the application of RoC on models that feature three Higgs doublets (see for instance \cite{Keus:2013hya,Kubo:2003iw,Gomez-Bock:2021uyu,Das:2014fea,Kuncinas:2020wrn,Kalinowski:2021lvw,CarcamoHernandez:2022vjk,BhupalDev:2017txh,Darvishi:2019ltl,Darvishi:2021txa}). All the above mentioned cases and extensions are the main subject of our future  work.

\section{Conclusions}

In the present work we described the application of the \textit{reduction of couplings} method on the Two Higgs Doublet Model.  In particular, the top Yukawa coupling and 
the quartic Higgs couplings are expressed in terms of the strong gauge coupling, treating the other two gauge couplings as corrections. 
The 1-loop reduction is performed at a boundary scale $M_{\text{\text{bdry}}}$, over which a covering theory is assumed. The demand for phenomenological viability of the model sets the scale over which new physics appear at $M_{\text{\text{bdry}}}\sim10^7$ GeV. The reduction gives five sets of solutions, which fit the experimental limits for the  top quark mass and the (light) Higgs boson mass, using as input only the gauge coupling values at $M_Z$, while it fixes the value of $\tan\beta\sim 2.2$.  Thus, the RoC method provides us with a powerful tool to reduce the number of free parameters of a given theory, guiding the direction of possible viable extensions of the Standard Model.

\subsection*{Acknowledgments}
We would like to thank Wojciech Kotlarski for pointing out inconsistencies in the RGEs of the initial manuscript. 
GP is supported by the Portuguese Funda\c{c}\~{a}o para a Ci\^{e}ncia e Tecnologia (FCT) under Contracts UIDB/00777/2020, and UIDP/00777/2020, these projects are partially funded through POCTI (FEDER), COMPETE, QREN, and the EU. GP has a postdoctoral fellowship in the framework of UIDP/00777/2020 with reference BL154/2022\_IST\_ID.
GP and GZ would like to thank CERN-TH for the hospitality and support. GZ would like to thank the MPP-Munich and DFG Exzellenzcluster 2181:STRUCTURES of Heidelberg University for support. MM acknowledge support from DGAPA-UNAM through PAPIIT project IN109321. MAMP acknowledges support from a CONACYT scholarship, and from partial scholarships from IF-UNAM project PRIDIF21-3 and PAPIIT project IN109321.

\bibliographystyle{JHEP-2}
\bibliography{main}

\end{document}